# Protein Evolution within a Structural Space

Eric J. Deeds\*, Nikolay V. Dokholyan†, Eugene I. Shakhnovich‡§

*\*Department of Molecular and Cellular Biology, Harvard University, 7 Divinity Avenue, Cambridge, MA 02138, USA*

*†Department of Biochemistry and Biophysics, University of North Carolina School of Medicine, Chapel Hill, NC 27599, USA*

*‡Department of Chemistry and Chemical Biology, Harvard University, 12 Oxford Street, Cambridge, MA 02138, USA*

§Corresponding Author: Eugene Shakhnovich
Department of Chemistry and Chemical Biology
Harvard University
Cambridge, MA 02138
Tel: 617-495-4130
Fax: 617-496-5948
Email: eugene@belok.harvard.edu

# ABSTRACT

Understanding of the evolutionary origins of protein structures represents a key component of the understanding of molecular evolution as a whole. Here we seek to elucidate how the features of an underlying protein structural "space" might impact protein structural evolution. We approach this question using lattice polymers as a completely characterized model of this space. We develop a measure of structural comparison of lattice structures that is analgous to the one used to understand structural similarities between real proteins. We use this measure of structural relatedness to create a graph of lattice structures and compare this graph (in which nodes are lattice structures and edges are defined using structural similarity) to the graph obtained for real protein structures. We find that the graph obtained from all compact lattice structures exhibits a distribution of structural neighbors per node consistent with a random graph. We also find that subgraphs of 3500 nodes chosen either at random or according to physical constraints also represent random graphs. We develop a divergent evolution model based on the lattice space which produces graphs that, within certain parameter regimes, recapitulate the scale-free behavior observed in similar graphs of real protein structures.

# INTRODUCTION

Understanding protein evolution has presented a problem of wide interest for many years (Dokholyan et al., 2002; Koonin et al., 2002; Ponting and Russell, 2002; Qian et al., 2001; Tatusov et al., 2001). In particular, the variety of protein structures observed has prompted the question of how these unique polymers might have evolved (Dokholyan et al., 2002; Karev et al., 2002; Koonin et al., 2002; Qian et al., 2001). Although an intriguing subject in its own right, the desire to understand protein structural evolution has been motivated in part by a desire to understand protein folding and function (Koonin et al., 2002).

The task of understanding protein structural evolution has relied on the analysis of structural similarities between proteins (Dokholyan et al., 2002; Karev et al., 2002; Koonin et al., 2002; Qian et al., 2001), much as the study of evolutionary relationships between genes and species has relied on studying similarities in primary sequence or other characters (Giribet, 2002). Structural similarity has been defined at varying levels of detail, from the assignment of structures to families and folds in human-annotated databases (Karev et al., 2002; Koonin et al., 2002; Qian et al., 2001) to the patterns of structural neighbors in quantitative comparisons (Dokholyan et al., 2002). Recently, graph theoretic approaches have been utilized to represent structural similarity at these varying scales, and have been used by many to motivate and implement various models of protein structural evolution (Dokholyan et al., 2002; Karev et al., 2002; Koonin et al., 2002; Qian et al., 2001). One particular application of these approaches, called the Protein Domain Universe Graph or PDUG (Dokholyan et al., 2002), revealed that the distribution of the number of structural neighbors $k$ per domain follows a power law $p(k)$

$\sim k^{-\mu}$ and represents a scale-free network (Albert and Barabasi, 2002; Barabasi and Albert, 1999).

The main result of the recent research in protein structural evolution has been the emergence of divergent evolution as a dominant paradigm (Dokholyan et al., 2002; Karev et al., 2002; Koonin et al., 2002; Qian et al., 2001). The divergent models that have been proposed to accompany these studies, while often successfully reproducing various global patterns of structural similarity observed in real proteins, have been nonetheless relatively abstract in nature. In particular, although the existence of a protein structural space or "universe" (that is, the set of all possible protein structures) has been postulated (Koonin et al., 2002), none of the proposed models have attempted to represent the evolution of proteins within this space (Dokholyan et al., 2002; Karev et al., 2002; Koonin et al., 2002; Qian et al., 2001). Indeed, there exists little understanding of what features (such as the underlying distribution of the number of structural neighbors) characterize this space, or how those features might influence structural evolution. Also, although the observed patterns of structural similarity have been explained in terms of evolution, it is unclear if those patterns of similarity are truly the imprint of evolutionary processes or simply a general feature of a space of compact polymer structures. A detailed characterization of the mechanisms of protein structural evolution will therefore not only require at least some level of understanding of the protein structures that have been available to organisms over the course of their evolution, but also an understanding of which patterns observed in real proteins likely contain information about their evolution and which might simply result from their identity as polymers.

An understanding of this space, however, is impeded by the fact that it is difficult to obtain a characterization of the space of real proteins. It is currently impossible to reliably predict the structure of a protein that shares no detectable sequence homology with a known structure (Bonneau et al., 2002; Lesk et al., 2001). Although structural genomics projects may soon provide a much greater structural description of the extant protein universe, the protein structures that have been discovered over the course of evolution may represent only a fraction of the structural possibilities available from protein structural space (Koonin et al., 2002). Thus a complete description of this space in terms of actual protein structures is infeasible at this time.

Model protein systems, such as lattice polymers, do not necessarily share these limitations. One can completely enumerate the structural space of 3x3x3 lattice polymers, for instance, subject only to the reasonable constraint that the polymers be maximally compact. The resulting 103346 distinct structures represent a completely characterized structural space. Although there are many important differences between this set of structures and real proteins (such as a lack of secondary structure elements), this space nonetheless presents an interesting and simple model system for study (Li et al., 1996; Mirny and Shakhnovich, 1996; Shakhnovich and Gutin, 1990). The cubic lattice polymer also represents a generalized compact polymer, and thus can provide the necessary "baseline" for what one would expect when looking at distributions of structural similarity in spaces of other compact polymers.

In order to understand the space of lattice polymers, we must develop a system for classifying and studying the relationships between lattice structures. Given the large number of structures, it is impractical to do so in a manner analogous to that used in the development of human-annotated databases such as SCOP and CATH (Murzin et al., 1995; Orengo et al., 2002). We thus utilize an automated method similar to that employed by FSSP (Holm and Sander, 1996): we develop a method of structural comparison between lattice polymers based on the contact maps of these polymers and produce a *Z*-score of structural similarity analogous to that produced by DALI (Dietmann and Holm, 2001; Holm and Sander, 1993). Given the success of graph theoretic approaches in the representation of real protein spaces (Dokholyan et al., 2002), we use this lattice *Z*-score to construct a graph in which the lattice structures are represented as nodes and structural similarity (above some *Z*-score cutoff) is used to define the edges between those nodes. This graph, termed the Lattice Structure Graph or LSG, is a completely characterized structural space that can be analyzed using the same methodologies that have been applied to the limited set of real protein structures available.

We also approach and characterize a number of features of certain subspaces of the LSG. We employ a recently developed measure of stability and "designability" of lattice structures (England and Shakhnovich, 2003) in order to determine the effects of physical constraints on this space. We also develop two modified versions of a "duplication and divergence" evolutionary model to ascertain if correctly chosen subsets of the LSG can recapitulate the scale-free features of the PDUG.

## RESULTS

### STRUCTURAL SIMILARITY ON THE LATTICE

The 3x3x3 cubic lattice polymers we consider can adopt 103346 distinct conformations corresponding to the complete set of self avoiding random walks on that lattice (Shakhnovich and Gutin, 1990). If two positions in the polymer are located next to each other in space but are not nearest neighbors in sequence, they are considered to be in contact. Each structure can thus be uniquely described by the matrix of contacts between its monomers. Working from this model, we apply a measure for structural similarity that calculates the amount of structural overlap between two conformations. This measure, $Q$, is the percentage of overlap between the contact matrices given by:

$$Q_{AB} = \frac{\sum_{i,j} \Delta_{ij}^A \Delta_{ij}^B}{N}, \tag{1}$$

where $\Delta_{ij}$ is 1 if positions i and j in the conformation are in contact and 0 otherwise, and $N$ is the total number of contacts possible in the polymer (in the case of the 3x3x3 cubic lattice, $N = 28$). The distribution of $Q$-scores between lattice structures is well fit by a Gaussian (data not shown), and thus we define the Z-score of the comparison:

$$Z_{AB} = \frac{Q_{AB} - \langle Q \rangle}{\sigma_Q}, \tag{2}$$

where $\langle Q \rangle$ is the average value of $Q_{AB}$ over all $AB$ pairs and $\sigma_Q$ is the standard deviation of $Q$ over all $AB$ pairs. In the case of the lattice, $\langle Q \rangle = 0.188$ and $\sigma_Q = 0.075$. This process produces a Z-score that is the exact analogue of the DALI Z-score (Dietmann and Holm, 2001) which may be used for structural comparisons in this set of conformations.

One can define a cutoff for structural similarity using the above framework in a manner similar to that employed for structural comparisons between real proteins. To assay the similarity between such comparisons on the lattice and in real proteins, we compared the distribution of Z-scores greater than a cutoff ($Z_{min}$) of 2 (the smallest Z-score reported by the DALI algorithm) on the lattice and in the set of protein domains that constitute the PDUG (Dokholyan et al., 2002).

In both cases, the distribution of Z-scores exhibit long, non-Gaussian tails (see Fig. 1). Although the scales of the Z-scores differ (given that there is a maximum Z of ~9 on the lattice), we observe that the distribution of structural similarity in this respect is quite similar—indicating that this most likely represents a feature of compact polymers under this framework of structural comparison.

**THE LATTICE STRUCTURE GRAPH**

Given the properties of this space outlined above, we constructed a graph using the lattice conformations and Z-scores in the same manner applied to create the PDUG (Dokholyan et al., 2002). In this case, we define individual lattice conformations as nodes, and define edges between those nodes based on whether their structural comparisons result in a Z-score above the $Z_{min}$ cutoff. The resulting LSG contains a unique set of structural relationships at each $Z_{min}$. Our analysis of this graph is performed at various values of this cutoff parameter.

One of the most telling features of any graph is its degree distribution, or the distribution of the number of edges (in this case structural neighbors) per node. We calculate the degree distribution at various values of $Z_{min}$ for the entire LSG, yielding the results in Fig. 2. In stark contrast to the power-law degree distribution observed in the PDUG (Dokholyan et al., 2002), we find that $p(k)$ for the LSG is well fit by a Gaussian. Gaussian degree distributions are a well-known feature of random graphs, or graphs in which edges are distributed randomly (Albert and Barabasi, 2002).

This finding indicates that the space of lattice structures, while presenting a similar pair wise distribution of structural Z-scores, does not share the scale-free nature of the PDUG. This indicates that the degree distribution observed in the PDUG is not simply a characteristic of generalized, compact polymers.  This finding also suggests that the complete space of compact polypeptides might also represent a random graph, and although the results are not necessarily transferable to the set of protein structures, this finding nonetheless raises the question of how an underlying random graph topology might influence protein structures and structural evolution.

Although the LSG does not share its degree distribution with the PDUG, this does not indicate that other features might not be observed within both systems.  Another striking feature exhibited by proteins is their distribution in structural "families"—certain structural classes (also known as "folds") contain many more representatives than other folds (Dokholyan et al., 2002; Finkelstein et al., 1993; Koonin et al., 2002; Qian et al., 2001).  Within our graph-theoretical framework, these folds correspond to disjoint clusters of nodes (Dokholyan et al., 2002), and the distribution of cluster sizes has been shown to follow a power law in various classification schemes (Dokholyan et al., 2002; Koonin et al., 2002; Qian et al., 2001).  Given that such distributions can also be observed in random graphs (Dokholyan et al., 2002), we desired to identify corresponding "families" of lattice structures.

The space of lattice structures is much too large to cluster using current computational resources, so in order to explore this question we consider smaller subgraphs of the LSG.  To facilitate direct comparison with the ~3500 nodes in the PDUG, we construct random subgraphs of the LSG with the same number of structures.  These subgraphs exhibit Gaussian degree distributions at various values of $Z_{min}$ (Fig. 3).  We cluster these subgraphs at various values of $Z_{min}$ (Dokholyan et al., 2002), and discover that, as observed in other random graphs(Dokholyan et al., 2002), they undergo a transition in the size of the largest cluster (or "giant component" of the graph)—similar to the percolation transition in this subspace (Albert and Barabasi, 2002; Cohen et al., 2002) (Fig. 4, *A*).

The degree distribution of the random subgraphs at the mid (critical) point in this transition ($Z_C = 5$) is that of a sparsely connected random graph (Fig. 3). The distribution of cluster sizes at this point is well-fit by a power law (see Fig. 4 *B*), as expected from the behavior of random graphs at the transition (Dokholyan et al., 2002). The average power-law exponent of the fits of these distributions is 2.3, roughly that observed when a similar approach is applied to both the PDUG and other fold-space representations, as well as random graphs derived from the PDUG (Dokholyan et al., 2002; Qian et al., 2001).

**PHYSICAL CONSTRAINTS**

Although the random-graph nature of the LSG may have implications for the nature of the real protein space, the preceding analysis entirely ignores the possible existence of physical constraints that might influence the characteristics of the space of "feasible" protein structures. Although the effects of such constraints can only be roughly represented on the lattice, it is possible to identify and model how they might affect a space of protein structures.

The two simplest constraints one can apply to real structures are that the native state should be sufficiently stable energetically (retain their fold in the presence of reasonable thermal fluctuations) and that the structures should be stable with respect to mutations (retain their fold in the presence of reasonable sequence fluctuations). Recently, a reliable measure of both of these properties was discovered: the trace of the contact matrix of a structure, which gives information about its "designability" (or the number of sequences that can adopt that structure) and its stability (England and Shakhnovich, 2003) (In Press). This measure correlates well with the stability and observed sequence entropy of lattice structures, and thus provides a good approximation to these physical constraints in this system.

We thus considered sets of 3500 lattice structures with the greatest, least, and median contact traces corresponding to the most, least, and median "designable" structures in the LSG. As with the random subgraphs, we cluster these graphs and identify the transition in the giant component of each (Fig. 5 *A*). We calculate the degree distributions and cluster size distributions for these subgraphs (Fig. 5, *B* and *C*). It is evident that the most designable structures, and to a lesser extent the least designable structures, represent a specific subset of the graph that is much more highly connected than one readily observes at random (Fig. 5 *C*). This behavior most likely results from the shared structural features that yield either exceptionally high or exceptionally low contact traces. These graphs, while certainly not a random subset of the space, nonetheless exhibit degree distributions well fit by Gaussian functions and are thus similar to those of random graphs.

These constraints do not reproduce the degree distribution observed in the PDUG. It is quite possible, however, that the designability of a given protein structure, and subsequently its thermal stability, may influence that structure's probability of discovery or "fixation" as a viable protein structure over the course of evolution. Although these effects can only be approximated for the lattice, they provide a potentially useful system for exploring how such preferences might influence structural evolution.

**EVOLUTIONARY MODEL**

The random-graph nature of the LSG, and the potentially similar nature of the space of real protein structures, raises the question of how evolution might proceed in such a structural space. To address this question, we created a version of our earlier model of evolution in an arbitrary space (Dokholyan et al., 2002) to model the evolution of lattice proteins. In order to preserve the "duplication and divergence" paradigm of our previous model (Dokholyan et al., 2002), we add "daughter" structures to the evolving subgraph according to their level of relatedness to a randomly chosen "parent" structure. At each duplication step, the newly added daughter node has some probability of diverging far enough to become an "orphan" (i.e. not connecting to the parent node) and some probability of being structurally related to the parent node.

The nature of the lattice space complicates the definition of "orphan" nodes, however. In our previous model the addition of an orphan is a completely local event— the new node does not connect the parent node and is not allowed to contact any other node on the graph. On the lattice, however, nodes that do not contact a given parent node at a particular $Z_{min}$ may still contact other nodes in the "evolved" subgraph. Thus the addition of "true" orphans (i.e. nodes with degree 0) requires a global understanding of the connectivity between a candidate orphan and any of the structures currently on the graph. Given the predominance of orphan nodes in the degree distributions of both the previous model and the PDUG (Dokholyan et al., 2002), we define orphans in our new model as nodes that have a degree of 0 in the evolving subgraph, rather than nodes that simply do not connect the chosen parent node. Orphan nodes are chosen with some probability $p_O$, analogous to the probability that a daughter will be chosen beyond the cutoff for similarity in our earlier model (Dokholyan et al., 2002).

Connected daughter structures, chosen with probability of 1-$p_O$, are defined as any node that connects the chosen parent node above a $Z_{min}$ cutoff. Although this is similar to the previous model (Dokholyan et al., 2002), in the first instantiation of this model we do not choose daughter nodes according to the level of similarity above the cutoff. The connectivity between non-orphan nodes that do not belong to a parent-daughter pair depends only on the underlying edges in the LSG. In this model, duplicate nodes are not allowed, resulting in an evolved subgraph consisting solely of unique lattice structures.

We find that this model is quite sensitive to two parameters: $p_O$ and the $Z_{min}$ at which edges are defined. Within certain regimes the model simply cannot create subgraphs of the target size (3500 nodes). The number of "potential orphans" in the template graph decreases exponentially as the algorithm progresses (data not shown), and the rate of this decrease depends on $p_O$ (the percentage of nodes that are added as orphans) and $Z_{min}$. At high $p_O$ and low $Z_{min}$ this decrease is too fast to allow for construction of a graph with 3500 nodes. Conversely, at low $p_O$ and high $Z_{min}$ the nodes added to the graph do not have enough available "daughters" in the template space to fulfill the requirements of the model. We do not observe a successful run of 3500 nodes outside of a very narrow range of the parameter space, and simulations that could be completed using any value of $p_O$ were only readily observed at a $Z_{min}$ of 4. The small size of this parameter regime may simply result from the restricted nature of the lattice space (the relatively small number of available nodes, for instance); nonetheless this behavior clearly illustrates the effects an underlying space may exert on structural evolution.

We cluster the resulting subgraphs at varying $Z_{min}$'s, and find that the $Z_C$ of the transition in the giant component depends strongly on the $p_O$ of the simulation (Fig. 6 *A*). The transitions at lower values of $p_O$ not only occur at higher $Z_{min}$'s but are also more gradual in nature. We analyze the degree distributions of graphs produced by this model at both a $Z_{min}$ of 4 (the $Z_{min}$ at which edges are defined during the evolutionary simulations) and at the $Z_C$ for each particular graph (Fig. 6, *C* and *D*) At a $Z_{min}$ of 4 the degree distributions of the resulting graphs are well fit by stretched exponential functions (Fig. 6 *C*), whereas at $Z_C$ one cannot make a distinction between stretched exponential and power-law fits due to the lack of data at higher values of *k*. Although not necessarily consistent with the degree distributions of random graphs, these evolved graphs do not exhibit the power-law degree distributions that characterize both the PDUG and the evolutionary graphs produced by our previous model (Dokholyan et al., 2002).

One feature of the PDUG that cannot be accurately represented by the above model is the existence of nearly identical structures (with DALI *Z*-scores > 20) that are nonetheless distinct due to the fact that they share no detectable sequence similarity (Fig. 1 *B*) (Dokholyan et al., 2002). Our requirement that each node on the subgraph be structurally distinct, along with the relatively limited structural repertoire of the LSG, prevents such situations from occurring. Similarly, the discrete nature of this space prevents the existence of very similar structures (as is possible in the continuous space represented by our previous model). To explore the influence of these features, we create a second model in which complete duplication events are allowed. Two identical lattice structures have a *Z*-score of ~10.8 (resulting from a *Q*-score of 1, see Eq. 2), representing the highly identical structures possible in both the PDUG and models developed for continuous, arbitrary spaces.

In the second model, orphan nodes are chosen in exactly the same manner as before, and thus the algorithm cannot proceed below a $Z_{min}$ of 4. The choice of connected daughters, however, proceeds quite differently. The new algorithm chooses a Z-score above the cutoff with equal probability from all Z-score bins (including Z greater than 10). A random node is then chosen from that class of similarity, regardless of whether or not it has already been placed in the subgraph. Complete duplication events occur when a daughter node is chosen from the $Z > 10$ bin, or when a node that already exists on the graph is chosen from a particular $4 < Z < 10$ bin. Not only does this model more faithfully reproduce the fact that extremely similar structures exist on the PDUG, it also allows for the choice of a greater number of nodes that are similar (if not identical) to a parent node (given the Z-score distribution in Fig. 1 *A*, less similar nodes are entropically favored in the first lattice evolution model proposed).

The transition in the giant component exhibited by the graphs produced under the second algorithm follows a similar pattern to the first, with $Z_C$ for the graphs at different values of $p_O$ similar between the two models (Fig. 6 *A* and Fig. 7 *A*). The degree distributions of graphs produced by this model at a $Z_{min}$ of 4 are also well fit by stretched exponential functions, although the "perfect" duplication events result in nodes of much higher connectivity in these graphs than in those produced by the previous model (compare Fig. 6 *C* and 7 *C*). At $Z_C$ for each respective $p_O$ we observe degree distributions that are well-fit by power-law functions (Fig. 7 *D*), indicating that subgraphs with strong scale-free character can be observed at the critical value of $Z_{min}$ in these graphs. The power-law exponent of the fit of the degree distribution for the $p_O = 0.1$ graph is -1.6, a value close to that observed for the PDUG and the graphs produced (in certain parameter regimes) by our original model (Dokholyan et al., 2002).

This finding demonstrates that scale-free subsets of a space of polymer structures can be obtained using rules motivated from the standpoint of divergent evolution. Thus, if the space of compact polypeptides does indeed represent a random graph similar to the LSG, it is clear that divergent evolutionary sampling of these structures by organisms has the potential to explain the scale-free nature of the PDUG.

## CONCLUSIONS

The LSG represents a particularly interesting system for the study of protein evolution. The random-graph nature of this system places the scale-free behavior of the PDUG in stark relief, indicating that the degree distribution of this graph may contain evolutionary information beyond the identity of proteins as polymer structures. Although a full understanding will require a complimentary analysis of real proteins, this analysis suggests that the space of possible protein structures represents a random graph as well. Random subgraphs of this space, and subgraphs chosen according to physical criteria, also display Gaussian degree distributions and thus do not recapitulate the behavior of the PDUG.

A striking feature of all of the subgraphs we analyze is the distribution of cluster sizes at the critical point. In every case, this distribution follows a power law, and the distributions are similar in terms of power-law exponent regardless of the rules used to generate each subgraph (compare Fig. 4 *B*, 5 *B*, 6 *B* and 7 *B*). This behavior seems to thus be a universal feature of these polymer subgraphs, and thus cannot be used to distinguish those that have been chosen according to evolutionary rules (Fig. 6 *B* and 7 *B*) or physical rules (Fig. 5 *B*) from those that have been chosen at random (Fig 4 *B*). Given that similar distributions have been reported for folds in real proteins (Qian et al., 2001), our lattice findings highlight the caution that must be employed when interpreting power law distributions of cluster or fold family sizes in terms of evolutionary processes.

We observe subgraphs with power-law *degree distributions* only as a result of a very specific evolutionary sampling procedure. This not only demonstrates that scale-free graphs may be derived from such spaces but also that the rules underlying divergent graph evolution models are sufficient to produce this behavior. Although the polymers we employ in this study are only rough approximations to proteins, the evolutionary graphs we produce consist of real structures and not of arbitrary nodes and edges. This data is thus not only useful as a proof of evolutionary principle but may also be seen as a source of structures with a more "realistic" distribution of structural similarity for other studies based on the 3x3x3 cubic lattice (Mirny and Shakhnovich, 1996).

One remaining challenge with respect to evolutionary models involves understanding the discovery of orphan structures in real biological systems. In our model we are capable of calculating the degree of candidate orphan nodes and can thus ensure that nodes of degree 0 will be added to the graph. This calculation is most likely not performed in similar kind by evolving organisms, and the question remains as to how a given fraction of orphan nodes will be discovered by over the course of protein evolution. Orphans may be a simple consequence of sequence dynamics, or they may result from a need for completely new structures to fulfill functional pressures. Further understanding of this phenomenon will require the development of more detailed models of structural evolution.

It is important to note that the conclusions we draw based on this model system, while quite rigorous due to the completeness of our description of this structural space, are not necessarily transferable to other systems. Although the features of the lattice space do provide a necessary control for structural comparisons involving general compact polymers, one must undertake further studies to asses the extent to which structural spaces that cannot be so fully understood exhibit similar behavior. In the case of polypeptides, it may be possible to use protein-like decoy structures (Bonneau et al., 2002; McConkey et al., 2003) in order to gain a glimpse of the features of the real protein space beyond those structures that have been crystallized and beyond the set of structures that have been discovered over the course of evolution. Nonetheless, it is clear that the lattice will continue to provide insights into protein structural evolution and the implications this evolution carries for the study of protein folding and protein function.

## ACKNOWLEDGEMENTS

We would like to thank Dr. B. N. Dominy, I. A. Hubner, J. E. Donald and E. Perlstein for their reading of the manuscript and J. L. England for his help. We thank the HHMI and the NIH for financial support. N.V.D. acknowledges support of The University of North Carolina at Chapel Hill Research Council grant.

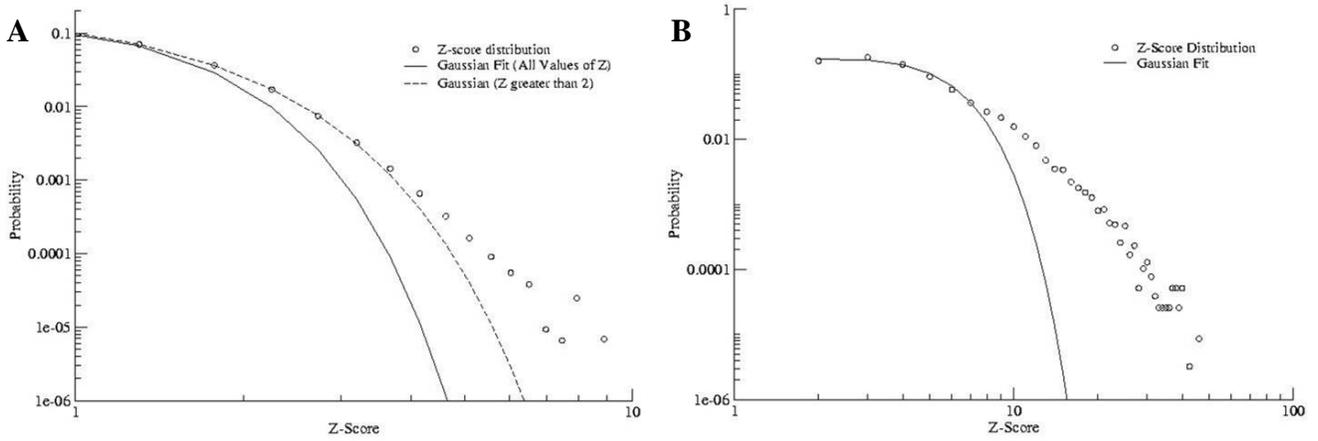

FIGURE 1  Z-Score distributions for (*A*) the LSG and (*B*) PDUG.  The Gaussian fits in *A* were performed using either the entire set (solid line) or the data for $Z > 2$ (for comparison with the fit in b).

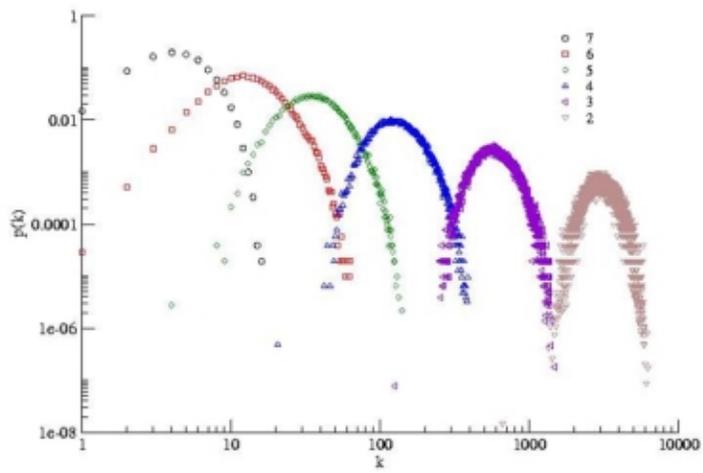

FIGURE 2 Degree distribution ($p(k)$) for the LSG at various values of $Z_{min}$. All degree distributions are shifted by a degree of one to allow display of degree 0 nodes on a log-log plot.

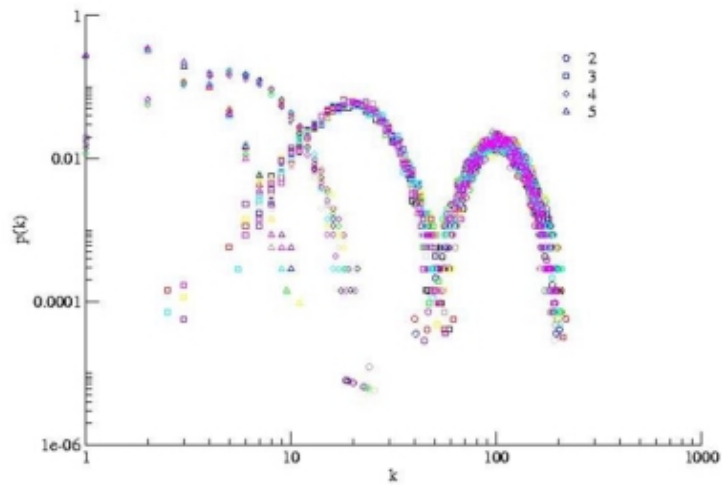

FIGURE 3 Degree distributions for 10 random subgraphs of the LSG containing 3500 nodes, at various values of $Z_{min}$. All degree distributions are shifted by a degree of one to allow display of degree 0 nodes on a log-log plot.

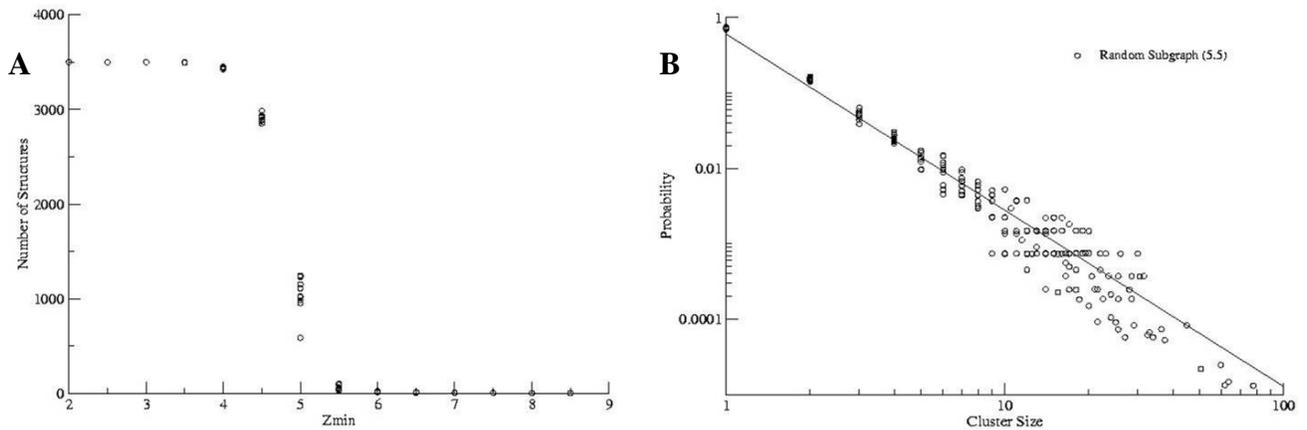

FIGURE 4  (*A*) Transition in the giant component of 10 random subgraphs of the LSG with 3500 nodes and (*B*) the cluster size distributions of those subgraphs at the $Z_C$ indicated in parentheses. The solid line in b represents a power-law regression of one of the random subgraphs. This particular regression exhibits a power-law exponent close to the average of -2.3.

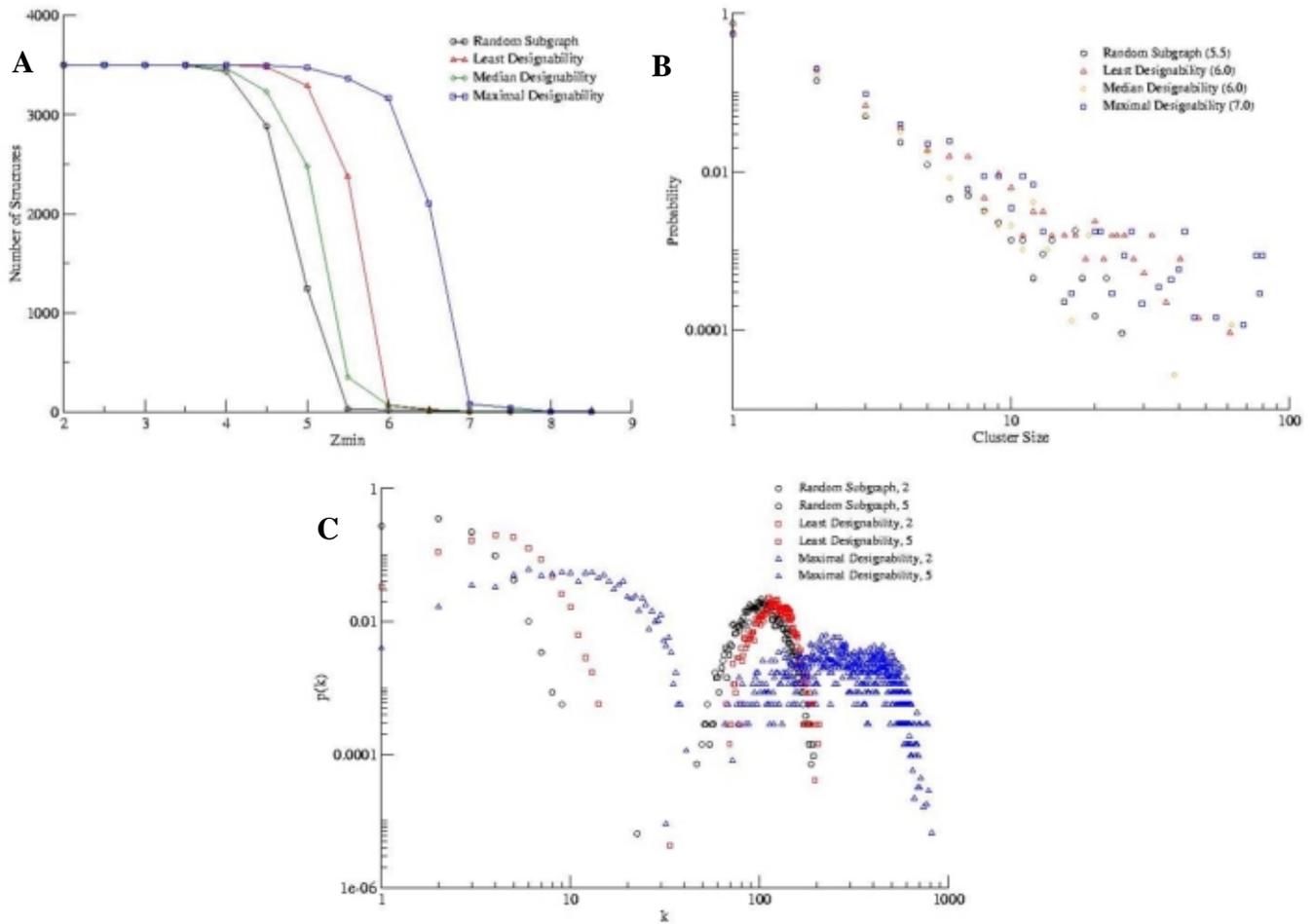

FIGURE 5 (*A*) Transition of the giant component subgraphs of 3500 nodes according to their contact trace (designability). (*B*) Cluster size distributions of these graphs at the $Z_{min}$ indicated in parentheses. (*C*) Degree distributions of these graphs at various values of $Z_{min}$. In each case an arbitrarily chosen random subgraph of 3500 nodes is included for comparison. All degree distributions are shifted by a degree of one to allow display of degree 0 nodes on a log-log plot.

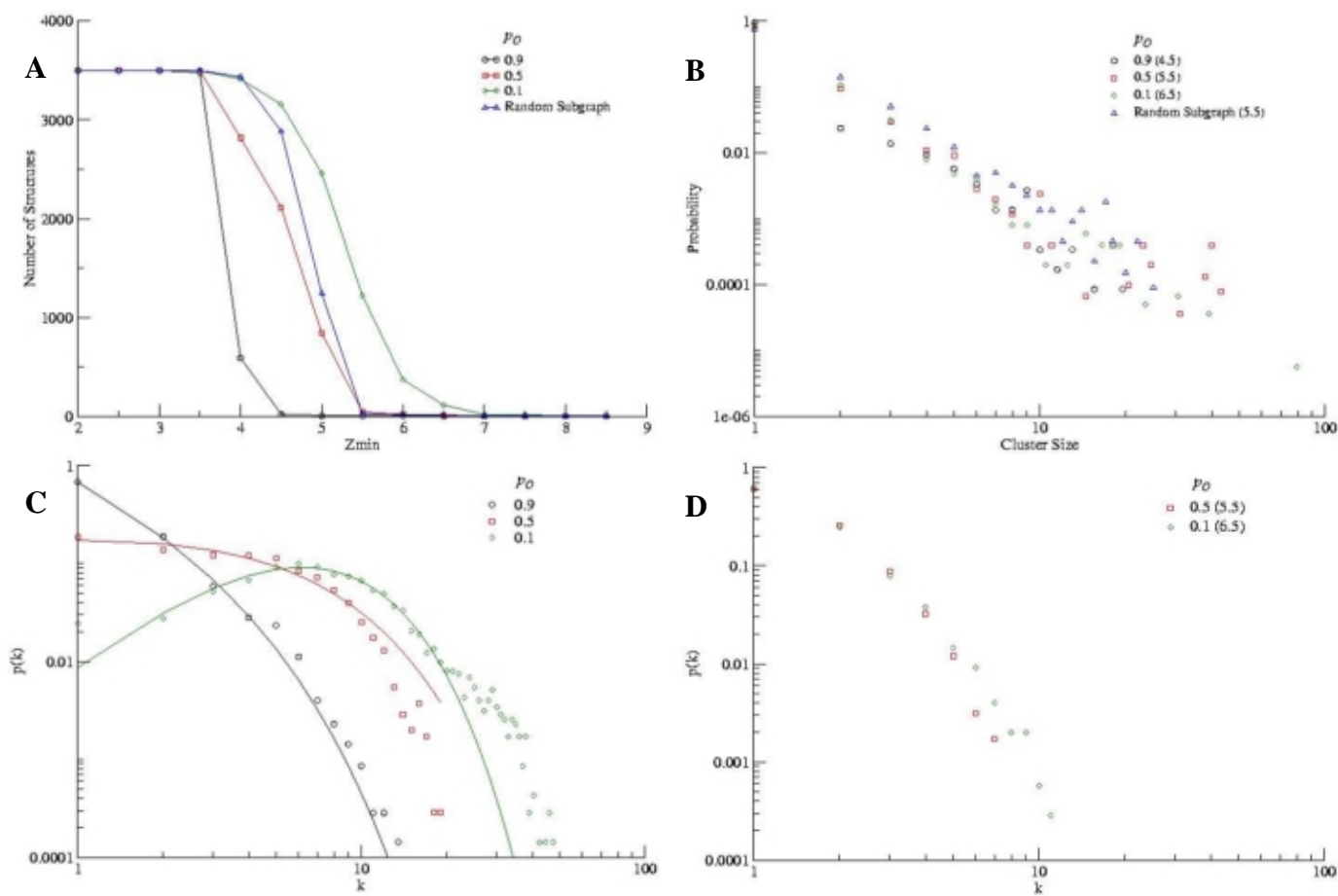

FIGURE 6 (*A*) Transition in the giant component and (*B*) cluster size distributions of evolved subgraphs produced using the first, non-duplication lattice evolution model. (*C*) Degree distributions of these graphs computed at a $Z_{min}$ of 4, the threshold at which the subgraphs were evolved. (*D*) The degree distributions in of evolved subgaphs calculated at the $Z_{min}$'s indicated in parentheses. For (*A*) and (*B*), a random subgraph is included for comparison. The solid lines in (*C*) represent stretched exponential fits. All degree distributions are shifted by a degree of one to allow display of degree 0 nodes on a log-log plot.

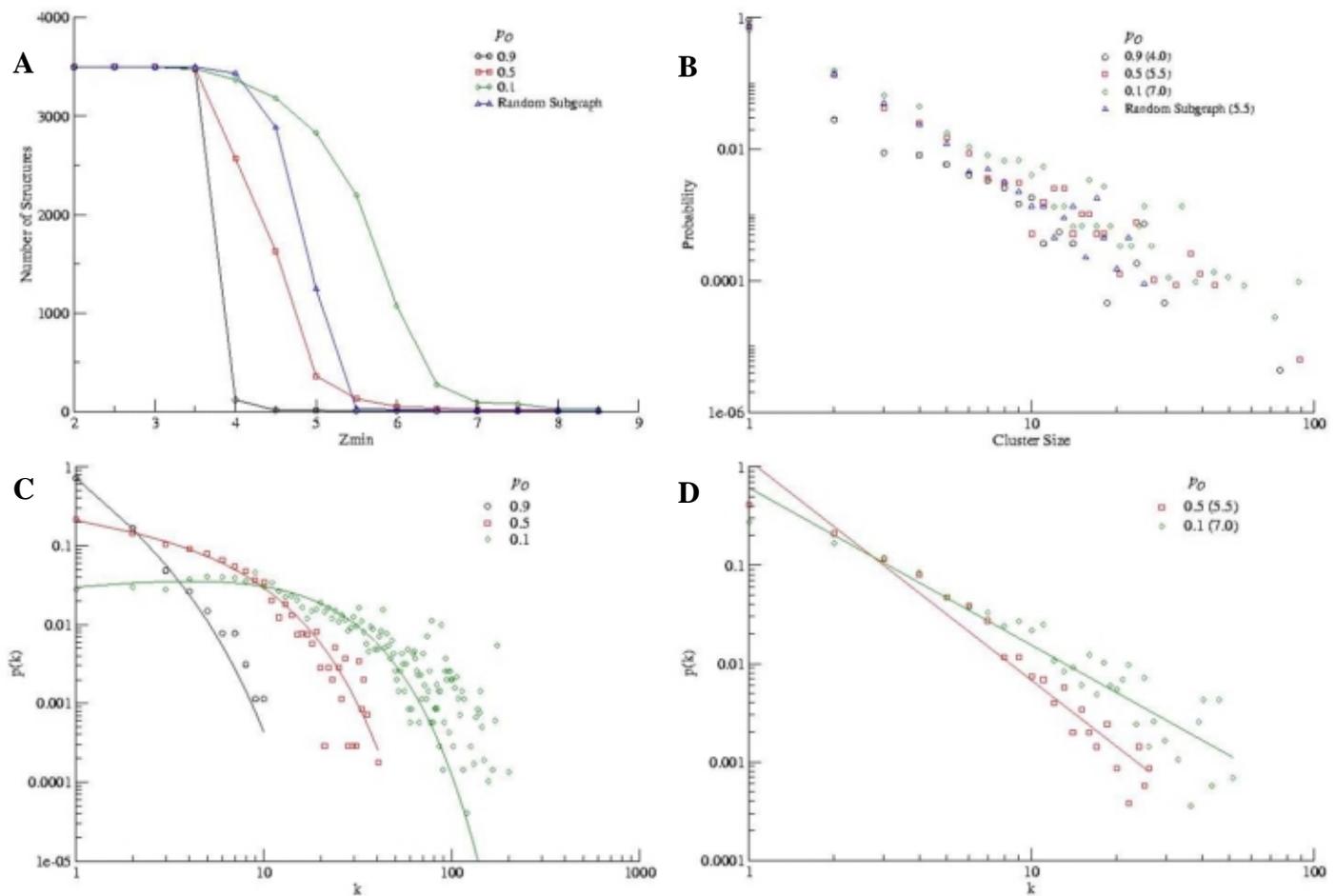

FIGURE 7 (*A*) Transition in the giant component and (*B*) cluster size distributions of evolved subgraphs produced using the second lattice evolution model in which duplication events are allowed. (*C*) Degree distributions of these graphs computed at a $Z_{min}$ of 4, the threshold at which the subgraphs were evolved. (*D*) The degree distributions in of evolved subgaphs calculated at the $Z_{min}$'s indicated in parentheses. For (*A*) and (*B*), a random subgraph is included for comparison. The solid lines in (*C*) represent stretched exponential fits; those in (*D*) represent power-law regressions. All degree distributions are shifted by a degree of one to allow display of degree 0 nodes on a log-log plot.